\documentclass[preprintnumbers,amsmath,11pt,amssymb,floatfix,superscriptaddress,nofootinbib]{article}

\def\mytitle#1{\setcounter{equation}{0}
\setcounter{footnote}{0}
\begin{flushleft}\Large\textbf{#1}\end{flushleft}
\vspace{0.20cm}}
\def\myname#1{\leftline{{\large #1}}\vspace{-0.13cm}}
\def\myplace#1#2{\small\begin{flushleft}\textit{#1}\\
\texttt{#2}\end{flushleft}}

\def\myclassification#1{\small\noindent
Keywords : Autonomous dynamical System, Dark energy, Dark matter, Interactions, Hyperbolic critical points, Phase Space .
       #1\vspace{0.5cm}}
\usepackage{graphicx}% Include figure files
\usepackage{wrapfig}
\usepackage{lipsum} % generates filler tex
\usepackage{multirow}
\usepackage[utf8]{inputenc}
\usepackage{array}
\usepackage{mathtools}% http://ctan.org/pkg/mathtools
\usepackage{makecell}
\usepackage{amsmath}
\usepackage{bigints}
\usepackage{amssymb}
\usepackage[mathscr]{euscript}
\usepackage{tipa}

\DeclareFontFamily{OT1}{pzc}{}
\DeclareFontShape{OT1}{pzc}{m}{it}{<-> s * [0.900] pzcmi7t}{}
\DeclareMathAlphabet{\mathpzc}{OT1}{pzc}{m}{it}
\begin{document}
\mytitle{Generalised Model of Interacting Dark Energy and Dark Matter : Phase Portrait Analysis of Evolving Universe}

\myname{ $Giridhari~Deogharia^{*}$\footnote{giridharideogharia@gmail.com}, $Mayukh~Bandyopadhyay^{*\dag}$\footnote{mayukhbandyopadhyay@yahoo.com}~and  $Ritabrata~
Biswas^{*}$\footnote{biswas.ritabrata@gmail.com}}
\myplace{*Department of Mathematics, The University of Burdwan, Golapbag Academic Complex, City: Burdwan-713104, District: Purba Barddhaman, State: West  Bengal, Country: India.\\$\dag$ Department of Physics, Bam Vivekananda P. T. T. College, Affiliated to The University of Burdwan, City: Burdwan-713101, District: Purba Barddhaman, State: West Bengal, Country: India.}{}
%%%%%%%%%%%%%%%%%%%%%%%%%%%%%%%%%%%%%%%%%%%%%%%%%%%%%%%%%%%%%%%%%%%%
\begin{abstract}
Main aim of this work is to give a suitable explanation of present accelerating universe through an acceptable interactive dynamical cosmological model. A three-fluid cosmological model is introduced in the background of Friedmann-Lemaître-Robertson-Walker asymptotically flat spacetime. This model consists of interactive dark matter and dark energy with baryonic matter taken as perfect fluid satisfying barotropic equation of state. We consider dust as the candidate of dark matter. A scalar field $\phi$ represents dark energy with potential $V(\phi)$. Einstein's field equations are utilised to construct a three-dimensional interactive autonomous system by choosing suitable interaction between dark energy and dark matter. We take the interaction kernel as $Q = 3\beta^{2\gamma} H\rho_d$. In order to explain the stability of this system, we obtained some suitable critical points. We analyse stability of obtained critical points to show the different phases of universe and cosmological implications. Surprisingly, we find some stable critical points which represent late time dark energy dominated era when a model parameter $\alpha=-5.05$. In order to explain both the energy dominated era as well as the late-time acceleration of the universe at same time, we introduce a two-dimensional interactive autonomous system. After graphical analysis of two-dimensional system, we get several stable points which represent dark energy dominated era and the late-time cosmic acceleration both at the same time. Here, we also shows the variation in interaction at vicinity of phantom barrier ($\omega_{eff}=-1$). From our work we can also predict the future phase evolution of the universe.

\end{abstract}
%%%%%%%%%%%%%%%%%%%%%%%%%%%%%%%%%%%%%%%%%%%%%%%%%%%%%%%%%%%%%%%%%%%%

\myclassification{\\PACS Numbers : 95.36.+x, 95.35.+d, 98.80.Cq}

%%%%%%%%%%%%%%%%%%%%%%%%%%%%%%%%%%%%%%%%%%%%%%%%%%%%%%%%%%%%%%%%%%%%%%%%%%%%%%%%%%%
\section{Introduction}
The exact shape and size of universe is a matter of debate. Experimental data from various numbers of recent independent observations, including type Ia supernovae (SNeIa) \cite{A. G. Riess,S. J. Perlmutter}, Large Scale Structure \cite{M. Tegmark,Percival}, Cosmic Microwave Background (CMB) \cite{D. N. Spergel,Komatsu} and Baryon  Acoustics  Oscillations (BAO) \cite{D. J. Eisenstein}  confirm that the universe is spatially flat with only a very small margin of error. Theoretical astrophysicists have been trying to construct a formal mathematical model of late-time universe. For this purpose, Friedmann-Lemaître-Robertson-Walker(hereafter FLRW) metric is mostly used. Expansion of universe is also well known fact and it has two accelerating phases. Firstly, a much slower and gradual expansion of space, about $ 10^{-32} $ of a second after the Big Bang which is the early acceleration phase (inflation)\cite{N. Mahata2015a}. This is generally known as the threshold of radiation dominated era. Second accelerating phase is a more recent: present time accelerated expansion. The later acceleration is faster. Cause of later faster accelerated expansion is ascribed to a mysterious exotic matter/energy with large negative pressure called dark energy(hereafter DE). The nature of DE can be described in different ways but none of these are fully understood. Idea that empty space can posses its own energy and Einstein's cosmological constant \cite{S. Weinberg,S. M. Carroll,P.J.E. Peebles}, together to a large extend, can explain accelerating phenomena barring two questions of cosmic coincidence problem and fine tuning problem. DE remains unclustered in all scales whereas in case of baryons and non-baryonic cold dark matter (hereafter DM), the same can be seen in the form of gravitational cluster. Recent cosmological observations and analysis claim that the energy budget of DE, DM and baryonic matter (hereafter BM) are $68\%$, $27\%$ and $5\%$ respectively \cite{N. Mahata2015a}.\\

In this article, we take into account F-L-R-W spacetime of universe. We consider universe is made up of three types of constituents- DE, DM and BM \cite{M. Tsamparlis}. DM is taken as pressureless dust. DE is described by quintessential scalar field and BM is hypothesised as a perfect fluid. Quintessential field can be considered to be a perfect fluid in effect, so the universe can be considered in terms of two perfect fluids and a dust \cite{N. Mahata2015a}.

After considering the conservation of mass and energy separately, we build our model. Considered type of interaction is minimally coupled to gravity. We also speculated that universe has got a dynamical stability under this  interaction. The evolution of energy density equations in case of DE and DM become
\begin{equation}\label{MBGD interaction}
  \dot{\rho}_{DM}+3H\rho_{DM}=Q~~~~~~and~~~~~~\dot{\rho}_{DE}+3H(1+\omega_{d})\rho_{DE}=-Q~~~~~~,
\end{equation}

where Q is taken as interaction kernel. This interaction can be denoted by different forms. The essential property of Q is that when $Q>0$, it indicates energy flow from DE to DM whereas $Q<0$ represents reverse flow of energy. Energy flow from DM to DE is not possible as it violates the generalised second law of thermodynamics \cite{N. Mahata2015b}. A constant $Q>0$ term is taken in the article like \cite{M. Cataldo}. Different models can be found in literature regarding interactions \cite{B. Wang,S. Wang,Y. L. Bolotin}. Another popular form of interaction is taken as $Q=H(\xi_{1}\rho_{DE}+\xi_{2}\rho_{DM})$. Reference \cite{M. Shahalam} has used the form $Q=\Gamma(\ddot{\rho}_{DE}+\ddot{\rho}_{DM})$. $Q=H\Gamma\rho_{DM}\rho_{DE}$ is other popular form mentioned in the reference \cite{C. Sergio}. In these cases $\rho_{DE}$ and $\rho_{DM}$ are energy densities of DE and DM respectively. $\Gamma$, $\xi_{1}$ and $\xi_{2}$ are parameters to describe the interaction strength in several works \cite{S. Wang,Y. Z. Ma}, applying holographic principle and discourses corresponding coupling quintessence model \cite{B. Richard}. Many researchers have taken Q in various ways in terms of energy densities $(\rho_{DM},\rho_{DE})$ and Hubble parameter, $H=\frac{\frac{da(t)}{dt}}{a(t)}$, $a(t)$ being the scale factor. In the reference \cite{C. van de Bruck,P. Biswas} the interaction kernel is taken as $Q=3Hc_{i}\rho_{i}$, where $i$=DM or DE, after considering the rate of energy transfer is proportional to the energy density of DE or DM and also proportional to $H$.

Several works on this category have already been done since the year of 2015. Some of them considered DE as complex scalar field \cite{R. Landim}. Again, In 2015, the work of Mahata and Chakraborty based on DBI DE model \cite{N. Mahata2015b} in order to explain the stability of a dynamical system. We plan to analyse the stability of the dynamical system in more general way and in our work we have derived the relation between interaction constant $\beta$ and energy densities of DE and DM.

Our investigation is focused on the study of the interaction given in equation (\ref{MBGD interaction}) and how universe has got stability under the same. We as well wish to explain the different phase evolution of universe and its future evolution also. Here we will develop an interacting autonomous dynamical system using Einstein's field equations and Klein-Gordon equation. We will study our model qualitatively. With the help of observational data and cosmological constraints, we will verify viable cosmological solutions. We will study stability of dynamical interactive autonomous system using acquired equilibrium points. Using feasible cosmological solution of our model we will try to depict different era of universe and also want to explain present percentages of DM and DE in universe and their evolutions. We will focus on the stability of equilibrium points obtained in order to explain the previous mentioned reason. We have followed the books \cite{L. Perko,D. K. Arrowsmith,S. Wiggins} to develop and explain interactive cosmological dynamical autonomous system.

The overview of our paper is as follows: in section 2, we will go through the mathematical modelling which contains two subsections. The first describes basic equations, i.e., Einstein's field equations, Klein-Gordon equation and energy conservation relations counting interactions between DE and DM. In the next subsection, we will constructed the three-dimensional interactive autonomous system considering some suitable variables. In section 3, we will analyse stability of our dynamical system considering interaction and some tables of critical points will be given. Section 4 describes reduced two-dimensional interactive autonomous system and contains some analysed phase portraits. Finally, section 5 sums up and ends up with brief discussions and important conclusions of this article.

\section{Mathematical Constructions}

\subsection{Basic Equations}

Here we assume DM in the form of dust with energy density $ \rho_{m} $ and choose a scalar field $ \phi $ with potential $ V(\phi) $ for DE.

The density $ \rho_{d} $ and pressure $ p_{d} $ of DE in terms of $ \phi $ and $ V(\phi) $ takes the form
 \begin{equation}\label{MBGD equation of pressure}
  p_{d} = \frac{1}{2}\dot{\phi}^2 - V(\phi)+ \frac{1}{2}I_{S}\beta^{r}\frac{3H^2}{k}~~~~~~and~~~~~~
  \end{equation}

  \begin{equation}\label{MBGD equation of density}
  ~\rho_{d} = \frac{1}{2}\dot{\phi}^2 + V(\phi)+ \frac{1}{2}I_{S}\beta^{r}\frac{3H^2}{k}~~~~~~.
 \end{equation}
We take real $ \phi $ for quintessence and $ I_S $, a suitable constant representing the change in scalar field $ \phi $ due to the interaction between DE and DM.
Here, the differentiation with respect to cosmic time is denoted by $ '\cdot' $.
Now, as the baryonic matter is considered like a perfect fluid, its equation of state in terms of pressure $ p_{b} $ and density $ \rho_{b} $ becomes
\begin{equation}\label{MBGD EoS}
  p_{b} = (\nu - 1)\rho_{b}~~~~~~,
\end{equation}
where $ \nu $ is the adiabatic index of baryonic fluid. Range of $ \nu $ is, $ \frac{2}{3}<\nu\leqslant 2 $ and for a particular case we get that $ \nu = 1 $ corresponds to DM and $ \nu = \frac{4}{3}$ fits with radiation dominated phenomena\cite{N. Mahata2015a}. In our model, we assume that DE and DM interact with each other and this interaction is minimally coupled to gravity.

In simple terms, the principle idea of general relativity is $geometry=k\times matter$, where $k$ is the coupling constant which ascertains the strength of gravitational force. Now, from Einstein's field equations, after assuming $ k= 8\pi G = 1 $, one of the Friedmann equation can be obtained as,
\begin{equation}\label{MBGD Friemann}
  3H^{2} = k(\rho_{m} + \rho_{d} + \rho_{b})~~~~~~.
 \end{equation}

For scalar field $ \phi $, the Klein-Gordon equation turns to be
\begin{equation}\label{MBGD Klein-Gordon}
  \ddot{\phi} + 3H\dot{\phi} + \frac{dV}{d\phi} = 0~.
\end{equation}

Energy conservation relations of our model after assuming the interaction between DE and DM, $ Q = 3\beta^{2\gamma} H\rho_d $ take the form,
\begin{equation}\label{MBGD DE}
  \dot{\rho_{d}} + 3H (\rho_{d} + p_{d}) = - Q = -3\beta^{2\gamma} H\rho_d~,
\end{equation}
\begin{equation}\label{MBGD DM}
  \dot{\rho_{m}} + 3H\rho_{m} = Q = 3\beta^{2\gamma} H\rho_d~~~and
\end{equation}
\begin{equation}\label{MBGD BM}
  \dot{\rho_{b}} + 3H (\rho_{b} + p_{b}) = 0~,
\end{equation}
where $ \beta $ is the interacting constant with $ \gamma = 1 $.
Now using the above equations from (\ref{MBGD Friemann}) to (\ref{MBGD BM}) we can derive,
\begin{equation}\label{MBGD Mixed}
   2\dot{H} = - k\left[\rho_{m}(1+I_{S}\beta^{r})+ \rho_{b}(\nu+I_{S}\beta^{r})+\dot{\phi}^2 + I_{S}\beta^{r}\rho_{d}\right]~~~~~~~~.
\end{equation}

Three main evolution equations of this model are the equations (\ref{MBGD Friemann}), (\ref{MBGD Klein-Gordon}) and (\ref{MBGD Mixed}) which are highly non-linear. In the following section, we use these equations to formulate the interacting dynamical system. We also construct suitable coordinate changes to make these equations compatible with the dynamical system.

\subsection{3D Interactive Autonomous System}

In our model, we introduce suitable coordinate transformations of dimensionless variables \cite{Amendola,E. J. Copeland} to form a dynamical system under interaction as,
\begin{equation}\label{MBGD parameter}
   x = \sqrt{\frac{k}{6}}\frac{\dot{\phi}}{H}~~~~~~ and ~~~~~~~
  y = \sqrt{\frac{k}{3}}\frac{\sqrt{V(\phi)}}{H}~~~~~.
\end{equation}

Now the dimensionless density parameters look like,
\begin{equation}\label{MBGD Density parameter}
\Omega_{m}=\frac{k\rho_{m}}{3H^{2}}~~~,~~~
\Omega_{b}=\frac{k\rho_{b}}{3H^{2}}
~~~~~and~~~~~
\Omega_{d}=\frac{k\rho_{d}}{3H^{2}}~~~~~.
\end{equation}

So, the Friedmann equations (\ref{MBGD Friemann}) and (\ref{MBGD Mixed}) transform into the following equations after the reasonable coordinate changes as given below,
\begin{equation}\label{MBGD D1}
  \Omega_{d}= x^{2}+y^{2}+\frac{1}{2}I_{S}\beta^{r}~~~~,~~~~
\end{equation}
\begin{equation}\label{MBGD D2}
\Omega_{d}= x^{2}+y^{2}+\frac{1}{2}I_{S}\beta^{r}~~~,~~~\Omega_{m}+\Omega_{b}+x^{2}+ y^{2}+\frac{1}{2}I_s\beta^{r} = 1~~~and
\end{equation}
\begin{equation}\label{MBGD H}
  \dot{H}= -\frac{3H^{2}}{2}\left[x^{2}(I_{S}\beta^{r}+2)+ \Omega_{m}(I_{S}\beta^{r}+1)+(\nu+I_{S}\beta^{r})\Omega_{b}+I_{S}\beta^{r}y^{2}+\frac{I_{S}\beta^{2r}}{2}\right].
\end{equation}

Here, $ \Omega_{m} $ and $ \Omega_{b} $ are non-negative real quantities representing the density parameters of DM and BM respectively. From equation (\ref{MBGD D1}), we get that, $ \Omega_{m} \leq 1 $ and $ 0 \leq \Omega_{b} \leq 1 $. So, $ x $ and $ y $ both satisfy $ x^2 + y^2+\frac{I_{S}\beta^{r}}{2} \leq 1 $ and it is clear that if energy densities related to DM and BM is zero then (\ref{MBGD D1}) becomes, $ x^2 + y^2+\frac{I_{S}\beta^{r}}{2} = 1 $.

Differentiating, equations (\ref{MBGD parameter}) and (\ref{MBGD Density parameter}) with respect to $ N $ where $ N =\ln\{ {a(t)}\} $ and using equations from (\ref{MBGD Klein-Gordon}) to (\ref{MBGD BM}), (\ref{MBGD D2}) and (\ref{MBGD H}), we can construct the interactive autonomous dynamical system given below

$$
\frac{dx}{dN}= \frac{3x}{2} \left[x^{2}({I_{S}\beta^{r}+2})-2+\Omega_{m}(I_{S}\beta^{r}+1)+I_S\beta^{r}y^2+\frac{1}{2}I_S\beta^{2r}\right.~~~~~~~~~~~~~~~~~~~~~~~~~~~~~~~~~~~~~~~~~~~~~~~~~~~~~~~~~~~~~~$$
\begin{equation}\label{MBGD 3D1}
 ~~~~~~~~~~~~~~~~~~~~~~~~~~~~~~~~~~~~~~~~~~~~~~~~~~~~~ \left.+(\nu+I_{S}\beta^{r})\left(1-\Omega_{m}-x^{2}-y^{2}-\frac{I_{S}\beta^{r}}{2}\right)\right]-\sqrt{\frac{3}{2k}}\frac{1}{V}\frac{dV}{d\phi}y^{2}
\end{equation}

 $$ \frac{dy}{dN} =
  \frac{y}{2} \left[2x \sqrt{\frac{3}{2k}}\frac{1}{V}\frac{dV}{d\phi} + 3x^2(I_{S}\beta^{r}+2)+ \frac{3}{2}I_{S}\beta^{2r}\right.~~~~~~~~~~~~~~~~~~~~~~~~~~~~~~~~~~~~~~~~~~~~~~~~~~~~~~~~~~~~~~~~~~~~~~~~~~~~~~~~
   $$
   \begin{equation}\label{MBGD 3D2}
     ~~~~~~~~~~~~~~~~~~~~~~\left.+3I_{S}\beta^{r}y^2+3\Omega_{m}(I_{S}\beta^{r}+1)+3(\nu + I_{S}\beta^{r})\left(1-\Omega_{m}-x^{2}-y^{2}-\frac{1}{2}I_{S}\beta^{r}\right)\right]~~~and
   \end{equation}

$$\frac{d\Omega_{m}}{dN}=3 \beta^{2}\left(x^{2}+y^{2}+\frac{I_{S}\beta^{r}}{2}\right)-3\Omega_{m}\bigg[1-\Omega_{m}(I_{S}\beta^{r}+1)-x^{2}(I_{S}\beta^{r}+2)~~~~~~~~~~~~~~~~~~~~~~~~~~~~~~~~~~~~~~~~~~~~~~~~~~~$$
\begin{equation}\label{MBGD 3D3}
  ~~~~~~~~~~~~~~~~~~~~~~~~~~~~~~~~~~~~~~~~~~~\left.+I_{S}\beta^{r}y^{2}+\frac{I_{S}\beta^{2r}}{2}-(\nu+I_{S}\beta^{r})\left(1-\Omega_{m}-x^{2}-y^{2}-\frac{1}{2}I_{S}\beta^{r}\right)\right].
\end{equation}

So, we can bring out cosmological parameters related to our interacting model using the above transformed variable as given below,
\begin{equation}\label{MBGD Cosmological constant1}
  \Omega_{d}= x^{2}+y^{2}+\frac{I_{S}\beta^{r}}{2}~,
\end{equation}
\begin{equation}\label{MBGD Cosmological constant2}
  \omega_{d}=\frac{p_{d}}{\rho_{d}}=\frac{x^{2}-y^{2}+\frac{I_{S}\beta^{r}}{2}}{x^{2}+y^{2}+\frac{I_{S}\beta^{r}}{2}}~,
\end{equation}
\begin{equation}\label{MBGD Cosmological constant3}
  \omega_{eff}=\frac{p_{d}+p_{b}}{\rho_{m}+\rho_{d}+\rho_{b}}=2x^{2}-1+\Omega_{m}+I_{S}\beta^{r}+\nu\left(1-\Omega_{m}-x^{2}-y^{2}-\frac{I_{S}\beta^{r}}{2}\right)~~~and
\end{equation}
\begin{equation}\label{MBGD q}
  q=-\left(1+\frac{\dot{H}}{H^{2}}\right)
=-\left[\left(1-\frac{3}{2}(x^{2}(I_{S}\beta^{r}+2)+\Omega_{m}(I_{S}\beta^{r}+1)+(\nu+I_{S}\beta^{r})\Omega_{b}+I_{S}\beta^{r}y^{2}+\frac{I_{S}\beta^{2r}}{2})\right)\right]~~~~~,
\end{equation}

where q is the asymptotic flatness parameter.

\section{Stability Analysis of Interactive Cosmological Model}

In this section, we are going to analyse stability of our interacting dynamical system representing by the equations (\ref{MBGD 3D1}) to (\ref{MBGD 3D3}). As it is assumed that $V(\phi)$ is exponential for considered phase of DE, so $\frac{1}{V}\frac{dV}{d\phi}=$constant. The value of this constant depends on the change in scalar field $\phi$ and as well as structure of the potential $V(\phi)$.

In references \cite{C. Wetterich}, time dependent scalar field, $\phi$ of quintessence is taken with its exponential potential $V(\phi)$ using Kaluza-Klein theories \cite{P. G. Ferreira}. This potential is also taken as  allowable model for quintessence in references \cite{A. R. Liddle}. It has a connection with inflation as it produces power-law expansion with some interesting properties \cite{F. Lucchin}.

Now as we take $$\sqrt{\frac{3}{2k}}\frac{1}{V}\frac{dV}{d\phi}=cosntant=\alpha(say)~~~~~,$$ the autonomous system equations from (\ref{MBGD 3D1}) to (\ref{MBGD 3D3}) transformed into the following forms:  \\
$$
\frac{dx}{dN}= \frac{3x}{2} \left[x^{2}({I_{S}\beta^{r}+2})-2+\Omega_{m}(I_{S}\beta^{r}+1)+I_S\beta^{r}y^2+\frac{1}{2}I_S\beta^{2r}\right.~~~~~~~~~~~~~~~~~~~~~~~~~~~~~~~~~~~~~~~~~~~~~~~~~~~~~~~~~~~~~~$$
\begin{equation}\label{MBGD S1}
 ~~~~~~~~~~~~~~~~~~~~~~~~~~~~~~~~~~~~~~~~~~~~~~~~~~~~~ \left.+(\nu+I_{S}\beta^{r})\left(1-\Omega_{m}-x^{2}-y^{2}-\frac{I_{S}\beta^{r}}{2}\right)\right]-\alpha y^{2}
\end{equation}

  $$ \frac{dy}{dN} =
  \frac{y}{2} \left[2\alpha x + 3x^2(I_{S}\beta^{r}+2)+ \frac{3}{2}I_{S}\beta^{2r}\right.~~~~~~~~~~~~~~~~~~~~~~~~~~~~~~~~~~~~~~~~~~~~~~~~~~~~~~~~~~~~~~~~~~~~~~~~~~~~~~~~~~~~~~~~~~~~~
   $$
   \begin{equation}\label{MBGD S2}
     ~~~~~~~~~~~~~~~~~~~~~~\left.+3I_{S}\beta^{r}y^2+3\Omega_{m}(I_{S}\beta^{r}+1)+3(\nu + I_{S}\beta^{r})\left(1-\Omega_{m}-x^{2}-y^{2}-\frac{1}{2}I_{S}\beta^{r}\right)\right]~~~and
   \end{equation}

$$\frac{d\Omega_{m}}{dN}=3 \beta^{2}\left(x^{2}+y^{2}+\frac{I_{S}\beta^{r}}{2}\right)-3\Omega_{m}\bigg[1-\Omega_{m}(I_{S}\beta^{r}+1)-x^{2}(I_{S}\beta^{r}+2)~~~~~~~~~~~~~~~~~~~~~~~~~~~~~~~~~~~~~~~~~~~~~~~~~~~$$
\begin{equation}\label{MBGD S3}
  ~~~~~~~~~~~~~~~~~~~~~~~~~~~~~~~~~~~~~~~~~\left.+I_{S}\beta^{r}y^{2}+\frac{I_{S}\beta^{2r}}{2}-(\nu+I_{S}\beta^{r})\left(1-\Omega_{m}-x^{2}-y^{2}-\frac{1}{2}I_{S}\beta^{r}\right)\right].
\end{equation}

For the present interactive model, we will obtain four sets of critical points, depending on the values of $\beta$, $\nu$ and $I_{S}$. These critical points are very effective to analyse the stability of this autonomous system. Using these critical points and stability criterions related to them, we can efficiently describe the different phase evolution of our universe from radiation dominated era to energy dominated era. The four sets of critical points are shown in Table 1:

\begin{center}
~~~~~~~~~~~~~~~~~~~~~~~~~~~~~~~~~~~ Table 1: Critical Points ~~~~~~~~~~~~~~~~~~~~~~~~~~~~~~~~~~~~
\end{center}
\begin{center}
\begin{tabular}{|l|p{16.5cm}|}
\hline
 C.P. &
Sets of Critical Points \\
\hline
$ C_{1} $  &  $\left(0,0,\frac{2(\nu-1)+\beta^{r}I_{S}\{2-\beta^{r}(1-I_{S})-\beta^{2-r}\nu\}-\sqrt{8\beta^{2+r}I_{S}(\nu-1)+[2(1-\nu)+\beta^{r}I_{S}\{\nu-2+\beta^{r}(1+I_{S})\}]^{2}}}{4(\nu-1)}\right)$ \\
\hline
$ C_{2} $  &  $\left(0,0,\frac{2(\nu-1)+\beta^{r}I_{S}\{2-\beta^{r}(1-I_{S})-\beta^{2-r}\nu\}+\sqrt{8\beta^{2+r}I_{S}(\nu-1)+[2(1-\nu)+\beta^{r}I_{S}\{\nu-2+\beta^{r}(1+I_{S})\}]^{2}}}{4(\nu-1)}\right)$  \\
\hline
$ C_{3} $  &  $\left(\frac{\sqrt{2(2-\nu)- \beta^{r}I_{S}(2+\nu)- I_{S}\beta^{2r}(5-2\nu+I_{S})+ \beta^{2+r}I_{S}(1-\nu)+ \beta^{3r}(I_{S})^2(2-\nu)+ \beta^{4r}(I_{S})^2(1-I_{S})}} {\sqrt{2(2-\nu)(1-\beta^2-I_{S}\beta^{2r})}},0,
 \frac{\beta^{2}\{4-\beta^{2r}I_{S}+\beta^{2r}(I_{S}-2\nu)^{2}\}}{2(2-\nu)(1-\beta^2-I_{S}\beta^{2r})}\right)$\\
\hline
$C_{4} $ & $\left(-\frac{\sqrt{2(2-\nu)- \beta^{r}I_{S}(2+\nu)- I_{S}\beta^{2r}(5-2\nu+I_{S})+ \beta^{2+r}I_{S}(1-\nu)+ \beta^{3r}(I_{S})^2(2-\nu)+ \beta^{4r}(I_{S})^2(1-I_{S})}} {\sqrt{2(2-\nu)(1-\beta^2-I_{S}\beta^{2r})}},0,
 \frac{\beta^{2}\{4-\beta^{2r}I_{S}+\beta^{2r}(I_{S}-2\nu)^{2}\}}{2(2-\nu)(1-\beta^2-I_{S}\beta^{2r})}\right)$\\
\hline

\end{tabular}
\end{center}

~~~~~~~~~~~~~~~~~~~~~~~~~~~~~~~~~~~~~~~~~~~~~~~~~~~~~~~~~~~~~~~~~~~~~~~~~~~~~~~~~~~~~~~~~~~~~~~~~~~~~~~~~~~~~~~~~~~~~~~~~~~~~~~~~~~~~~~~~~~~~~~~~~~~~~~~~~~~~~~~~
~~~~~~~~~~~~~~~~~~~~~~~~~~~~~~~~~~~~~~~~~~~~~~~~~~~~~~~~~~~~~~~~~~~~~~~~~~~~~~~~~~~~~~~~~~~~~~~~~~~~~~~~~~~~~~~~~~~~~~~~~~~~~~~~~~~~~~~~~~~~~~~~~~~~~~~~~~~~~~~~~

Now, using the sets of critical points $ C_{1} $ and $ C_{2} $ given in Table 1 and with the help of suitable values of $ I_{S} $, $ \beta $ and $ \nu $, we first calculate the values of dimensionless density parameters $ \Omega_{m} $, $ \Omega_{d} $ and  $ \Omega_{b} $ with $r=1$.

 Table 2 represents the early phase evolution of the universe. In Table 2, sets of critical points $ C_{1} $ and $ C_{2} $ show the variation in interaction between DE and DM. We also get their stable or unstable conditions from the eigen values given in the Table 2.
\\
\\
\\
\\

\begin{center}
 Table 2\\
 First \& Second sets of critical points: Radiation and matter dominated era
\end{center}
\begin{center}
\begin{tabular}{|l|l|l|l|l|l|p{2cm}|p{2cm}|p{2cm}|p{3cm}|}
\hline
C.P. & $x$ & $y$ & $ \nu $ & $ I_{S} $ & $ \beta $ & $ \Omega_{m} $ & $\Omega_{d}$ & $\Omega_{b}$ & Eigen Values \\
\hline
$C_{1a}$ & 0 & 0 & 1.2 & 0 & 0 & 0 & 0 & 1 & 1.80, -1.20, 0.60 \\
\hline
$C_{2a}$ & 0 & 0 & 1.6 & 0 & 0 & 1 & 0 & 0 & -1.80, -1.5, 1.5 \\
\hline
$C_{1b}$ & 0 & 0 & 0.9 & 0.5 & 0.001 & 0.9973 & 0.0002 & 0.0025 & 1.50, -1.50, 0.30 \\
\hline
$C_{1c}$ & 0 & 0 & 0.85 & 0.4 & 0.009 & 0.984 & 0.004 & 0.012 & 1.50, -1.49, 0.44 \\
\hline
$C_{1d}$ & 0 & 0 & 0.675 & 1.1 & 0.01 & 0.978 & 0.006 & 0.016 & 1.50, -1.49, 0.95 \\
\hline
$C_{1e}$ & 0 & 0 & 0.685 & 0.7 & 0.08 & 0.895 & 0.028 & 0.077 & 1.51, -1.49, 0.84 \\
\hline
$C_{1f}$ & 0 & 0 & 0.69 & 1.2 & 0.1 & 0.639 & 0.120 & 0.241 & 1.52, -1.48, 0.73 \\
\hline
$C_{1g}$ & 0 & 0 & 0.68 & 2.6 & 0.28 & 0.465 & 0.378 & 0.157 & 1.71, -1.29, 0.23 \\
\hline
$C_{1h}$ & 0 & 0 & 0.71 & 2.62 & 0.29 & 0.440 & 0.380 & 0.180 & 1.72, -1.28, 0.169 \\
\hline
$C_{1i}$ & 0 & 0 & 0.70 & 2.55 & 0.3 & 0.438 & 0.383 & 0.179 & 1.73, -1.27, 0.16 \\
\hline
$C_{1j}$ & 0 & 0 & 0.71 & 2.51 & 0.31 & 0.426 & 0.389 & 0.185 & 1.73, -1.26, 0.11 \\
\hline
$C_{1k}$ & 0 & 0 & 0.73 & 2.51 & 0.32 & 0.413 & 0.402 & 0.185 & 1.74, -1.26, 0.04 \\
\hline
$C_{2b}$ & 0 & 0 & 0.72 & 2.33 & 0.37 & 0.402 & 0.431 & 0.167 & 1.76, -1.24, -0.10 \\
\hline
$C_{2c}$ & 0 & 0 & 1.1 & 2.63 & 0.382 & 0.398 & 0.502 & 0.100 & 1.79, -1.20, -0.67 \\
\hline
$C_{2d}$ & 0 & 0 & 0.9 & 2.55 & 0.4 & 0.359 & 0.510 & 0.131 & 1.77, -1.23, -0.57 \\
\hline
$C_{2e}$ & 0 & 0 & 1.25 & 2.65 & 0.45 & 0.331 & 0.596 & 0.073 & 1.76, -1.34, -1.24 \\
\hline
$C_{2f}$ & 0 & 0 & 1.5 & 2.73 & 0.46 & 0.315 & 0.628 & 0.057 & -1.74, 1.73, -1.27 \\
\hline

\end{tabular}
\end{center}

~~~~~~~~~~~~~~~~~~~~~~~~~~~~~~~~~~~~~~~~~~~~~~~~~~~~~~~~~~~~~~~~~~~~~~~~~~~~~~~~~~~~~~~~~~~~~~~~~~~~~~~~~~~~~~~~~~~~~~~~~~~~~~~~~~~~~~~~~~~~~~~~~~~~~~~~~~~~~~~~~~
~~~~~~~~~~~~~~~~~~~~~~~~~~~~~~~~~~~~~~~~~~~~~~~~~~~~~~~~~~~~~~~~~~~~~~~~~~~~~~~~~~~~~~~~~~~~~~~~~~~~~~~~~~~~~~~~~~~~~~~~~~~~~~~~~~~~~~~~~~~~~~~~~~~~~~~~~~~~~~~~~~

Third and forth sets of critical points are calculated on the basis of real scalar field $ \phi $. Here, $x$ is taken real and positive. Table 3 represents third set of critical points which are obtained by the same procedure as for the critical points of Table 2. We also have the forth set of critical points which are not shown as they are same as third set of critical points.

\begin{center}
 Table 3\\
 Third set of critical points: DE dominated era
\end{center}
\begin{center}
\begin{tabular}{|l|l|l|l|l|l|l|p{2cm}|p{2cm}|p{2cm}|p{3cm}|}
\hline
C.P. & x & y & $ \nu $ & $ I_{S} $ & $ \beta $ & $ r $ & $ \Omega_{m} $ & $\Omega_{d}$ & $\Omega_{b}$ & Eigen Values \\
\hline
$C_{3a}$ & 0.5975 & 0 & 1.99 & 0.530 & 1.3 & 1 & 0.202 & 0.702 & 0.096 & -0.12+2.32i, -0.12-2.32i, 0.59(5.04+$ \alpha $) \\
\hline
$C_{3b}$ & 0.5692 & 0 & 1.975 & 0.450 & 1.35 & 1 & 0.205 & 0.628 & 0.167 & 0.018+2.28i, 0.018-2.28i, 0.57(5.31+$ \alpha $) \\
\hline
$C_{3c}$ & 0.5980 & 0 & 1.99 & 0.515 & 1.38 & 1 & 0.226 & 0.713 & 0.061 & -0.27+2.44i, -0.27-2.44i, 0.60(5.04+$ \alpha $) \\
\hline
$C_{3d}$ & 0.60146 & 0 & 1.98 & 0.520 & 1.4 & 1 & 0.229 & 0.726 & 0.045 & -0.32+2.48i, -0.32-2.48i, 0.60(5.02+$ \alpha $) \\
\hline
$C_{3e}$ & 0.60647 & 0 & 1.97 & 0.523 & 1.45 & 1 & 0.239 & 0.747 & 0.014 & -0.43+2.55i, -0.43-2.55i, 0.61(5.0+$ \alpha $) \\
\hline
$C_{3f}$ & 0.5963 & 0 & 1.972 & 0.49 & 1.5 & 1 & 0.268 & 0.723 & 0.016 & -0.47+2.6i, -0.47-2.6i, 0.60(5.08+$ \alpha $) \\
\hline

\end{tabular}
\end{center}

~~~~~~~~~~~~~~~~~~~~~~~~~~~~~~~~~~~~~~~~~~~~~~~~~~~~~~~~~~~~~~~~~~~~~~~~~~~~~~~~~~~~~~~~~~~~~~~~~~~~~~~~~~~~~~~~~~~~~~~~~~~~~~~~~~~~~~~~~~~~~~~~~~~~~~~~~~~~~~~~~~~
~~~~~~~~~~~~~~~~~~~~~~~~~~~~~~~~~~~~~~~~~~~~~~~~~~~~~~~~~~~~~~~~~~~~~~~~~~~~~~~~~~~~~~~~~~~~~~~~~~~~~~~~~~~~~~~~~~~~~~~~~~~~~~~~~~~~~~~~~~~~~~~~~~~~~~~~~~~~~~~~~~~
In Table 2, we have obtained the critical points to explain radiation dominated era as well as matter dominated era of universe whereas the critical points of Table 3 represents the energy dominated era of universe. Naturally, critical points of Table 2 are unstable but some of the critical points of Table 3 are stable in certain conditions. After the critical point $C_{2f}$ in Table 2, we got the points which represents energy dominated era with $x=y=0$. We did not want to take $x=y=0$ for energy dominated era of universe. Thus, Table 3 signifies the critical points with $x\neq0$ and $y=0$.

Here, in order to explain the different era of universe, using stability analysis of critical points of interactive autonomous system chosen, we have got some hyperbolic critical points with eigen values which are related to cosmological parameters. In Table 2, critical point $ C_{1a} $ represents entire domination of the non-accelerating universe by BM when there is no such interaction took place between DE and DM. This unstable critical point illustrates the radiation dominated phase of the universe. The next critical point $ C_{2a} $ signifies the non-accelerating universe entirely dominated by DM which is known as matter dominated era of universe. Here, no interaction between DE and DM has occurred. This critical point is also unstable. Now from other critical points $ C_{1b} $ to $ C_{1k} $, we can get idea of expansion of universe at its early stage when the universe is in matter dominated era. Unstable critical points $ C_{2b} $ to $ C_{2f}$ define the start of DE domination of the accelerated phase of universe. It is also clear from the values of $ \Omega_{d} $ and the unstable critical points, $ C_{1b} $ to $ C_{2f} $ in Table 2 that interaction between DE and DM increases due to increase in amount of DE. In this case, at the beginning, there is a small existence of DE which is continuously increasing with the continuous decrease in DM.

Now, if we look at the critical points of Table 3 and assumed the value of $\alpha= -5.05$, the stable critical point $C_{3a}$ denotes almost full domination of the DE which actually represents present phase evolution of universe. The critical point $C_{3b}$ has a special significance, because the interaction between DE and DM increases but the value of $\Omega_d$ decreases which opposes our assumption and this is why for any value of $\alpha$ it remains unstable. After assuming the value of $\alpha=-5.05$, we get that critical points $C_{3c}$ to $C_{3e}$ are all stable. These points signify the energy dominated accelerated expansion of the universe and show the increasing interaction between DM and DE of present situation. Interaction increases due to the increase in the value of $\Omega_d$.

For critical point $C_{3f}$, if we take $\alpha=-5.05$, this point becomes unstable. From the values of density parameters and interaction between DE and DM related with this critical point, we can imagine about the probable change in the values of $\Omega_d$ and $\Omega_m$ in future. It may be seen that the value of $\Omega_m$ will further increase whereas the value of $\Omega_d$ will decrease if there is more further interactions between DE and DM. In Table 3, from the critical points from $C_{3a}$ to $C_{3f}$, it is clear that in the DE dominated era of universe as interaction between DM and DE increases, value of $\Omega_m$ also increases. But, using this three-dimensional interactive cosmological model we can not explain DE dominated era and late-time acceleration of universe both at same time.

\section{2D Interacting Autonomous System}

In this section, we made a 2D interacting autonomous system, after neglecting the evolution of DM. This is a reduced system from our above 3D interacting autonomous system. Therefore, our 2D autonomous system becomes:

$$
\frac{dx}{dN}= \frac{3x}{2}\left[x^{2}({I_{S}\beta^{r}+2})-2+\Omega_{m}(I_{S}\beta^{r}+1)+I_S\beta^{r}y^2+ \right.~~~~~~~~~~~~~~~~~~~~~~~~~~~~~~~~~~~~~~~~~~~~~~~~~~~~~~~~~~~~~~$$
\begin{equation}\label{MBGD 2D1}
  ~~~~~~~~~~~~~~~~~~~~~~~~~~~~~~~~~~~~~~~~~~~~~~~\left.\frac{1}{2}I_S\beta^{2r}+(\nu+I_{S}\beta^{r})\left(1-\Omega_{m}-x^{2}-y^{2}-\frac{I_{S}\beta^{r}}{2}\right)\right]-\alpha y^{2}~~~~and~~~
\end{equation}

 $$ \frac{dy}{dN} = \frac{y}{2}\left[ 2x\alpha + 3x^2(I_{S}\beta^{r}+2)+ \frac{3}{2}I_{S}\beta^{2r}+ 3I_{S}\beta^{r}y^2+3\Omega_{m}(I_{S}\beta^{r}+1)\right.~~~~~~~~~~~~~~~~~~~~~~~~~~~~~~~~~~~~~~~~~~~~$$
\begin{equation}\label{MBGD 2D2}
  ~~~~~~~~~~~~~~~~~~~~~~~~~~~~~~~~~~~~~~~~~~~~~~~~~~~~~~~~~~~~~~\left. +3(\nu + I_{S}\beta^{r})\left(1-\Omega_{m}-x^{2}-y^{2}-\frac{1}{2}I_{S}\beta^{r}\right)\right]
\end{equation}

Here, the recent study on DE dominated universe predicts that in present time value of $ \Omega_{m} $ is approximately 0.23. We vary the value of $ \Omega_{m} $ such that $0.19\leq \Omega_{m} \leq 0.32$.

For this system, clearly $(0,0)$ is a critical point. To evaluate the other critical points taking $y=0$ then we get $x=\pm\frac{\sqrt{-4+2\beta^{r} I_{S}+\beta^{2r} I_{S}-\beta^{2r} (I_{S})^{2}+2\Omega_{m}+2\nu-\beta^{r} I_{S}\nu-2\Omega_{m}\nu}}{\sqrt{2}\sqrt{\nu-2}} $. Again, for non-zero $(x,y)$ as critical points we have taken help of graphs. For different values of $ \beta $, $ \nu $, $ \alpha $, $ I_{S} $, $ \Omega_{m} $ and $ r=1 $, we get different critical points. With the help of these points graphs (1) to (12) are drawn and we enlist obtained critical points in tables.

\begin{center}
 Table 4 : Critical Points When $ \Omega_{m}=0.20 $, $ I_{S}=0.10 $, $ \beta=0.20 $, $ \alpha=1.51 $ and $ \nu=1.8 $
\begin{tabular}{|p{2.5cm}|l|l|l|l|l|l|p{4cm}|}
\hline
Critical Points & x & y & $ \Omega_{d} $ & $ \Omega_{b} $ & $ \omega_{eff} $ & $q$ &  Behaviour\\
\hline
$ C_{5a} $ & 0 & 0 & 0.01 & 0.79 & 0.563 & 1.4657 & Saddle Point \\
\hline
$ C_{5b} $ & -0.44 & 0.77 & 0.79 & 0.01 & -0.38 & -0.077  & Stable Node \\
\hline
$ C_{5c} $ & -0.44 & -0.77 & 0.79 & 0.01 & -0.38 & -0.077 & Stable Node \\
\hline
$ C_{5d} $ & 1.33 & 0 & 1.78 & Undefined & 1 & 1.99 & Unstable Node \\
\hline
$ C_{5e} $ & -1.33 & 0 & 1.78 & Undefined & 1 & 1.99 & Unstable Node \\
\hline

\end{tabular}
\end{center}

~~~~~~~~~~~~~~~~~~~~~~~~~~~~~~~~~~~~~~~~~~~~~~~~~~~~~~~~~~~~~~~~~~~~~~~~~~~~~~~

\begin{center}
~~~~~~~~~~~~~~~~~ Table 5: Critical Points when $ \Omega_{m}=0.23 $, $ I_{S}=0.11 $, $ \beta=0.28 $, $ \alpha=1.6 $, $ \nu=1.6 $ ~~~~~~~~~~~~~~~
\begin{tabular}{|p{2.5cm}|l|l|l|l|l|l|p{4cm}|}
\hline
Critical Points & x & y & $ \Omega_{d} $ & $ \Omega_{b} $ & $ \omega_{eff} $ & $ q $ &  Behaviour\\
\hline
$ C_{6a} $ & 0 & 0 & 0.0154 & 0.7546 & 0.47 & 1.21 & Saddle Point \\
\hline
$ C_{6b} $ & -0.42 & 0.76 & 0.7694 & 0.0006 & -0.39 & -0.072 & Stable Node \\
\hline
$ C_{6c} $ & -0.43 & -0.75 & 0.7628 & 0.0072 & -0.36 & -0.072 & Stable Node \\
\hline
$ C_{6d} $ & 1.15 & 0 & 1.3379 & Undefined & 0.99716 & 2 & Unstable Node\\
\hline
$ C_{6e} $ & -1.15 & 0 & 1.3379 & Undefined & 0.99716 & 2 & Unstable Node\\
\hline

\end{tabular}
\end{center}

\begin{center}
 Table 6: Critical Points when  $ \Omega_{m}=0.22 $, $ I_{S}=0.2 $, $ \beta=0.3 $, $ \alpha=1.1 $, $ \nu=1.55 $
\centering
\begin{tabular}{|p{2.5cm}|l|l|l|l|l|l|p{4cm}|}
\hline
Critical Points & x & y & $ \Omega_{d} $ & $ \Omega_{b} $ & $ \omega_{eff} $ & $q$ &  Behaviour\\
\hline
$ C_{7a} $ & 0 & 0 & 0.03 & 0.75 & 0.4425 & 1.17 & Saddle Point \\
\hline
$ C_{7b} $ & -0.33 & 0.8 & 0.7789 & 0.0011 & -0.5 & -0.24 & Stable Node \\
\hline
$ C_{7c} $ & -0.33 & -0.8 & 0.7789 & 0.0011 & -0.5 & -0.24 & Stable Node \\
\hline
$ C_{7d} $ & 1.1 & 0 & 1.24 & Undefined & 0.987 & 1.99 & Unstable Node \\
\hline
$ C_{7e} $ & -1.1 & 0 & 1.24 & Undefined & 0.987 & 1.99 & Unstable Node \\
\hline

\end{tabular}
\end{center}

\begin{center}
 Table 7: Critical Points when  $ \Omega_{m}=0.24 $, $ I_{S}=0.22 $, $ \beta=0.35 $, $ \alpha=1 $, $ \nu=1.35 $

\begin{tabular}{|p{2.5cm}|l|l|l|l|l|l|p{4cm}|}
\hline
Critical Points & x & y & $ \Omega_{d} $ & $ \Omega_{b} $ & $ \omega_{eff} $ & $q$  & Behaviour\\
\hline
$ C_{8a} $ & 0 & 0 & 0.039 & 0.721 & 0.291025 & 0.95 & Saddle Point \\
\hline
$ C_{8b} $ & -0.31 & 0.79 & 0.7587 & 0.0013 & -0.489045 & -0.22 & Stable Node \\
\hline
$ C_{8c} $ & -0.31 & -0.79 & 0.7587 & 0.0013 & -0.489045 & -0.22 & Stable Node \\
\hline
$ C_{8d} $ & 1.0366 & 0 & 1.11 & Undefined & 0.989476 & 1.99 & Unstable Node\\
\hline
$ C_{8e} $ & -1.0366 & 0 & 1.11 & Undefined & 0.989476 & 1.99 & Unstable Node\\
\hline

\end{tabular}
\end{center}

\begin{center}
 Table 8: Critical Points when  $ \Omega_{m}=0.28 $, $ I_{S}=0.23 $, $ \beta=0.32 $, $ \alpha=1.2 $, $ \nu=1.99 $

\begin{tabular}{|p{2.5cm}|l|l|l|l|l|l|p{4cm}|}
\hline
Critical Points & x & y & $ \Omega_{d} $ & $ \Omega_{b} $ & $ \omega_{eff} $ & $q$ &  Behaviour\\
\hline
$ C_{9a} $ & 0 & 0 & 0.037 & 0.683 & 0.713168 & 1.58 & Saddle Point \\
\hline
$ C_{9b} $ & -0.30 & 0.77 & 0.7197 & 0.0003 & -0.465803 & -0.19 & Stable Node \\
\hline
$ C_{9c} $ & -0.30 & -0.77 & 0.7197 & 0.683 & -0.465803 & -0.19 & Stable Node \\
\hline

\end{tabular}
\end{center}

\begin{center}
 Table 9: Critical Points when $ \Omega_{m}=0.30 $, $ I_{S}=0.24 $, $ \beta=0.38 $, $ \alpha=1.15 $, $ \nu=1.7 $

\begin{tabular}{|p{2.5cm}|l|l|l|l|l|l|p{4cm}|}
\hline
Critical Points & x & y & $ \Omega_{d} $ & $ \Omega_{b} $ & $ \omega_{eff} $ & $q$ &  Behaviour\\
\hline
$ C_{10a} $ & 0 & 0 & 0.0456 & 0.6544 & 0.50368 & 1.27 & Saddle Point \\
\hline
$ C_{10b} $ & -0.29 & 0.75 & 0.6922 & 0.078  & -0.42734 & -0.28 & Stable Node \\
\hline
$ C_{10c} $ & -0.29 & -0.75 & 0.6922 & 0.078 & -0.42734 & -0.28 & Stable Node \\
\hline
$ C_{10d} $ & 1.27 & 0 & 1.6332 & Undefined & 0.97996 & 2 & Unstable Node\\
\hline
$ C_{10e} $ & -1.27 & 0 & 1.6332 & Undefined & 0.97996 & 2 & Unstable Node\\
\hline

\end{tabular}
\end{center}

\begin{center}
 Table 10: Critical Points when $ \Omega_{m}=0.32 $, $ I_{S}=0.5 $, $ \beta=-1.1 $, $ \alpha=1.6 $, $ \nu=1.55 $

\begin{tabular}{|p{2.5cm}|l|l|l|l|l|l|p{4cm}|}
\hline
Critical Points & x & y & $ \Omega_{d} $ & $ \Omega_{b} $ & $ \omega_{eff} $ & $q$ &  Behaviour\\
\hline
$ C_{11a} $ & -0.42 & 0.88 & 0.6758 & 0.0042  & -0.87069 & -0.57916 & Stable Node \\
\hline
$ C_{11b} $ & -0.42 & -0.88 & 0.6758 & 0.0042 & -0.87069 & -0.57916 & Stable Node \\
\hline
$ C_{11c} $ & -1.114 & 0 & 0.97 & Undefined & 0.808698 & 1.94 & Unstable Node\\
\hline
$ C_{11d} $ & 1.114 & 0 & 0.97 & Undefined & 0.808698 & 1.94 & Unstable Node\\
\hline

\end{tabular}
\end{center}

\begin{center}
Table 11: Critical Points when $ \Omega_{m}=0.27 $, $ I_{S}=0.54 $, $ \beta=-0.9 $, $ \alpha=1.5 $, $ \nu=1.8 $

\begin{tabular}{|p{2.5cm}|l|l|l|l|l|l|p{4cm}|}
\hline
Critical Points & x & y & $ \Omega_{d} $ & $ \Omega_{b} $ & $ \omega_{eff} $ & $q$ &  Behaviour\\
\hline
$ C_{12a} $ & -0.39 & 0.91 & 0.72 & 0.01 & -0.93 & -0.74 & Stable Node \\
\hline
$ C_{12b} $ & -0.39 & -0.91 & 0.72 & 0.01 & -0.93 & -0.74 & Stable Node\\
\hline
$ C_{12c} $ & -1.395 & 0 & 1.69 & Undefined & 0.92 & 2.03 & Unstable Node\\
\hline
$ C_{12d} $ & 1.395 & 0 & 1.69 & Undefined & 0.92 & 2.03 & Unstable Node\\
\hline

\end{tabular}
\end{center}

\begin{center}
Table 12: Critical Points when $ \Omega_{m}=0.23 $, $ I_{S}=0.75 $, $ \beta=-0.7 $, $ \alpha=1.2 $, $ \nu=1.1 $

\begin{tabular}{|p{2.5cm}|l|l|l|l|l|l|p{4cm}|}
\hline
Critical Points & x & y & $ \Omega_{d} $ & $ \Omega_{b} $ & $ \omega_{eff} $ & $q$ &  Behaviour\\
\hline
$ C_{13a} $ & -0.36 & 0.92 & 0.7135 & 0.0565 & -0.97 & -0.89 & Stable Node \\
\hline
$ C_{13b} $ & -0.36 & -0.91 & 0.7135 & 0.0565 & -0.97 & -0.89 & Stable Node\\
\hline
$ C_{13c} $ & -1.109 & 0 & 0.9674 & Undefined & 0.95 & 1.99 & Unstable Node\\
\hline
$ C_{13d} $ & 1.109 & 0 & 0.9674 & Undefined & 0.95 & 1.99 & Unstable Node\\
\hline

\end{tabular}
\end{center}

\begin{center}
Table 13: Critical Points when $ \Omega_{m}=0.22 $, $ I_{S}=1.05 $, $ \beta=-0.63 $, $ \alpha=1.7 $, $ \nu=1.4 $

\begin{tabular}{|p{2.5cm}|l|l|l|l|l|l|p{4cm}|}
\hline
Critical Points & x & y & $ \Omega_{d} $ & $ \Omega_{b} $ & $ \omega_{eff} $ & $q$ &  Behaviour\\
\hline
$ C_{14a} $ & -0.42 & 0.93 & 0.71055 & 0.0694 & -0.99147 & -1.003 & Stable Node \\
\hline
$ C_{14b} $ & -0.42 & -0.93 & 0.71055 & 0.0694 & -0.99147 & -1.003 & Stable Node\\
\hline
$ C_{14c} $ & -1.226 & 0 & 1.17 & Undefined & 1.02 & 2.007 & Unstable Node\\
\hline
$ C_{14d} $ & 1.226 & 0 & 1.17 & Undefined & 1.02 & 2.007 & Unstable Node\\
\hline

\end{tabular}
\end{center}

\begin{center}
Table 14: Critical Points when $ \Omega_{m}=0.20 $, $ I_{S}=1.09 $, $ \beta=-0.4 $, $ \alpha=1.33 $, $ \nu=1.49 $

\begin{tabular}{|p{2.5cm}|l|l|l|l|l|l|p{4cm}|}
\hline
Critical Points & x & y & $ \Omega_{d} $ & $ \Omega_{b} $ & $ \omega_{eff} $ & $q$ &  Behaviour\\
\hline
$ C_{15a} $ & -0.33 & 0.95 & 0.7934 & 0.0066 & -1.00837 & -1.02432 & Stable Node \\
\hline
$ C_{15b} $ & -0.33 & -0.95 & 0.7934 & 0.0066 & -1.00837 & -1.02432 & Stable Node\\
\hline
$ C_{15c} $ & -1.236 & 0 & 1.3097 & Undefined & 1.05994 & 2.07815 & Unstable Node\\
\hline
$ C_{15c} $ & 1.236 & 0 & 1.3097 & Undefined & 1.05994 & 2.07815 & Unstable Node\\
\hline

\end{tabular}
\end{center}

\begin{center}
Table 15: Critical Points when $ \Omega_{m}=0.19 $, $ I_{S}=1.12 $, $ \beta=-0.37 $, $ \alpha=1.36 $, $ \nu=1.2 $

\begin{tabular}{|p{2.5cm}|l|l|l|l|l|l|p{4cm}|}
\hline
Critical Points & x & y & $ \Omega_{d} $ & $ \Omega_{b} $ & $ \omega_{eff} $ & $q$ &  Behaviour\\
\hline
$ C_{16a} $ & -0.31 & 0.95 & 0.7914 & 0.0186 & -1.00988 & -1.02862 & Stable Node \\
\hline
$ C_{16b} $ & -0.31 & -0.95 & 0.7914 & 0.0186 & -1.00988 & -1.02862 & Stable Node \\
\hline
$ C_{16c} $ & -1.124 & 0 & 1.05618 & Undefined & 1.00694 & 1.99661 & Unstable Node\\
\hline
$ C_{16c} $ & 1.124 & 0 & 1.05618 & Undefined & 1.00694 & 1.99661 & Unstable Node\\
\hline

\end{tabular}
\end{center}

\begin{center}
$~~~~~~~~~~~~~~~~~~Fig.-1 ~~~~~~~~~~~~~~~~~~~~~~~~~~~~~~~~~~~~~~~~~~~~~~~~~~~~~~~~~~~Fig.-2~~~~~~~~~~~~~~~~~~~$\\
\includegraphics[height=1.7in,width=3in]{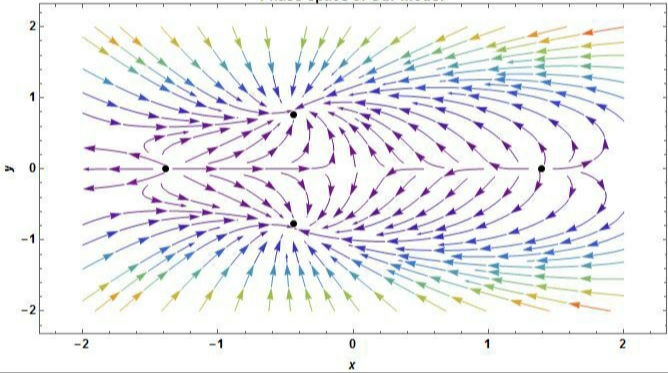}~~~~\includegraphics[height=1.7in,width=3in]{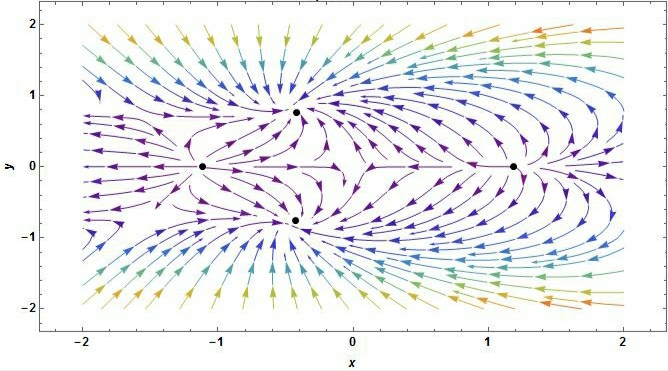} \\
$~~~~~~~~~~~~~~~~~~Fig.-3 ~~~~~~~~~~~~~~~~~~~~~~~~~~~~~~~~~~~~~~~~~~~~~~~~~~~~~~~~~~~Fig.-4~~~~~~~~~~~~~~~~~~~$\\
\includegraphics[height=1.7in,width=3in]{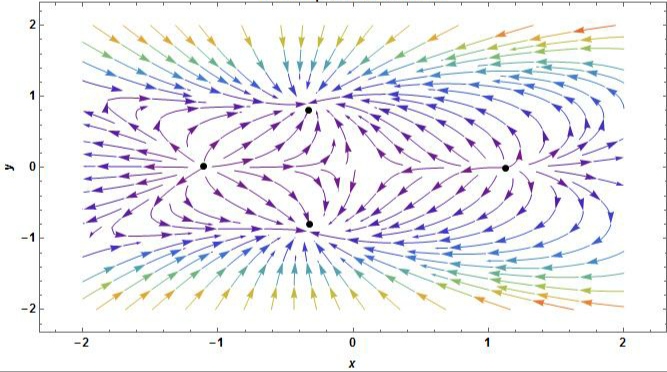}~~~~\includegraphics[height=1.7in,width=3in]{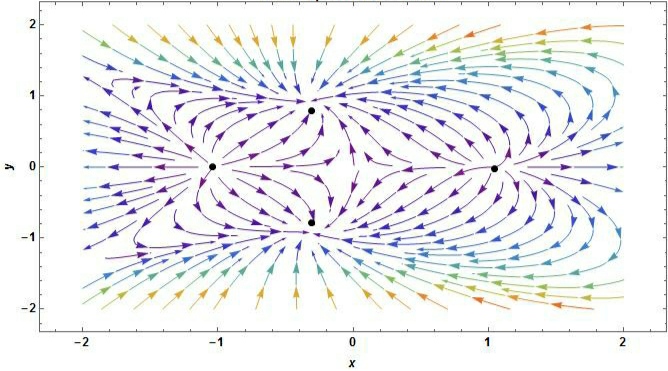} \\
$~~~~~~~~~~~~~~~~~~Fig.-5 ~~~~~~~~~~~~~~~~~~~~~~~~~~~~~~~~~~~~~~~~~~~~~~~~~~~~~~~~~~~Fig.-6~~~~~~~~~~~~~~~~~~~$\\
\includegraphics[height=1.7in,width=3in]{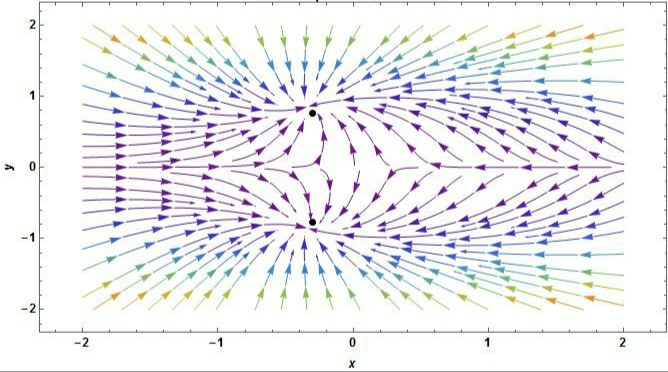}~~~~\includegraphics[height=1.7in,width=3in]{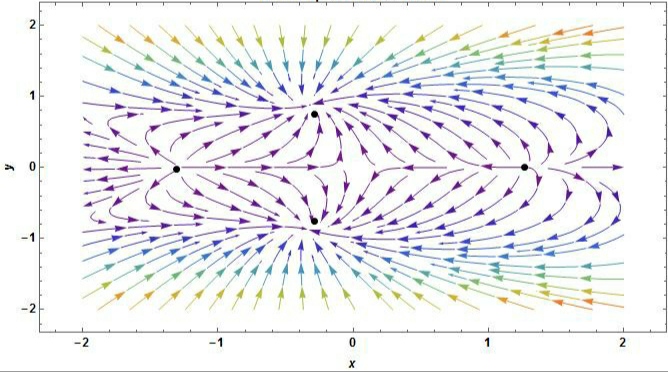} \\
$~~~~~~~~~~~~~~~~~~Fig.-7 ~~~~~~~~~~~~~~~~~~~~~~~~~~~~~~~~~~~~~~~~~~~~~~~~~~~~~~~~~~~Fig.-8~~~~~~~~~~~~~~~~~~~$\\
\includegraphics[height=1.7in,width=3in]{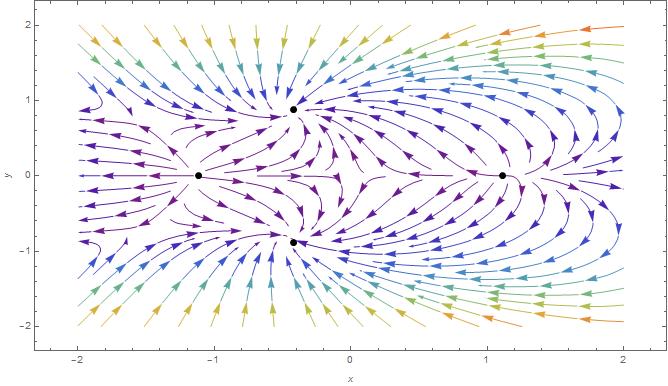}~~~~\includegraphics[height=1.7in,width=3in]{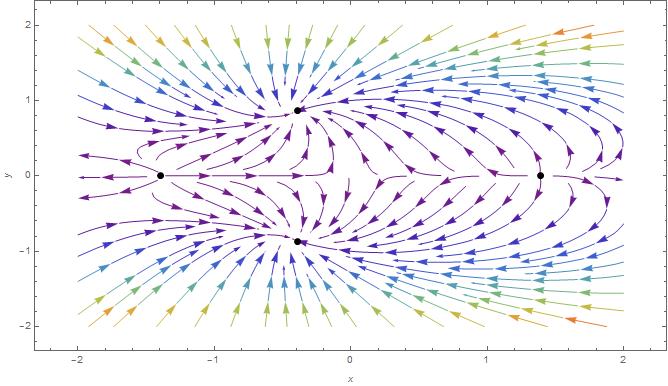} \\
$~~~~~~~~~~~~~~~~~~Fig.-9 ~~~~~~~~~~~~~~~~~~~~~~~~~~~~~~~~~~~~~~~~~~~~~~~~~~~~~~~~~~~Fig.-10~~~~~~~~~~~~~~~~~~~$\\
\includegraphics[height=1.7in,width=3in]{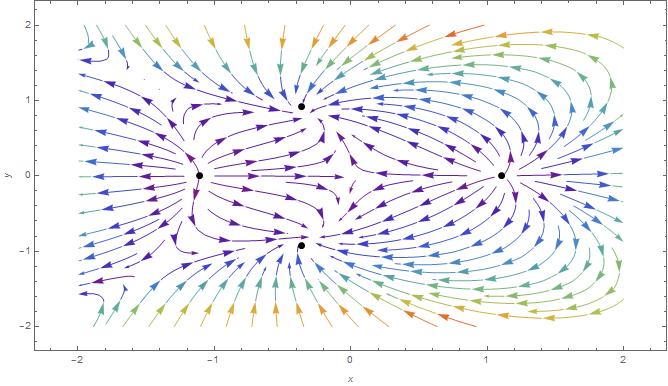}~~~~\includegraphics[height=1.7in,width=3in]{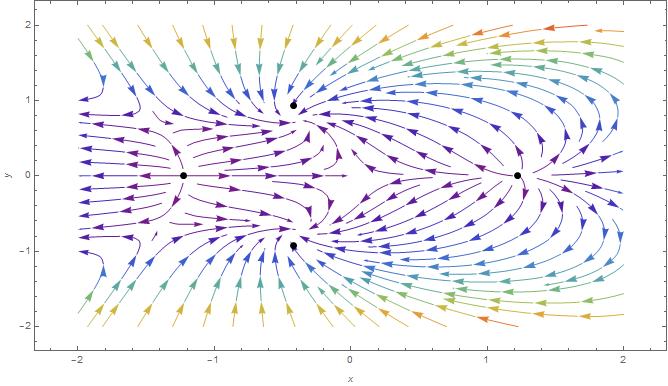} \\
$~~~~~~~~~~~~~~~~~~Fig.-11 ~~~~~~~~~~~~~~~~~~~~~~~~~~~~~~~~~~~~~~~~~~~~~~~~~~~~~~~~~~~Fig.-12~~~~~~~~~~~~~~~~~~~$\\
\includegraphics[height=1.7in,width=3in]{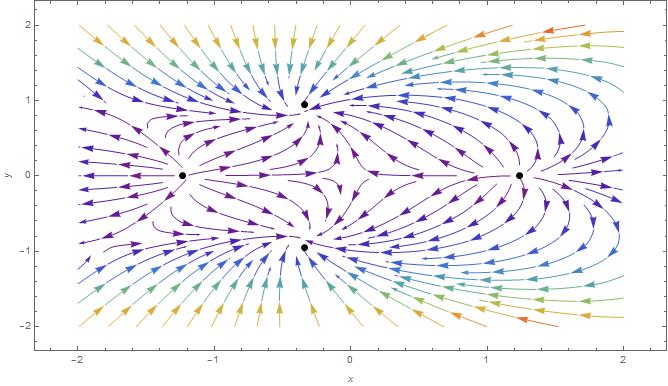}~~~~\includegraphics[height=1.7in,width=3in]{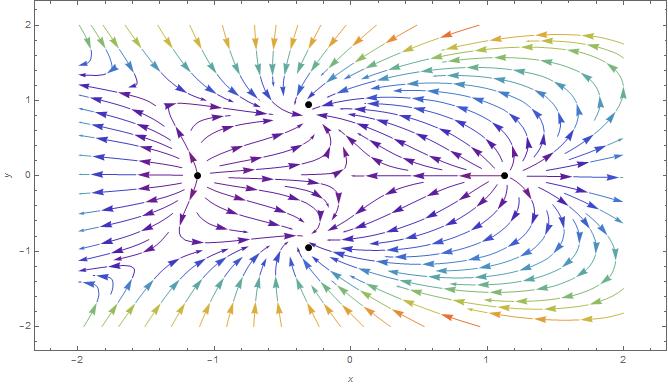} \\
Fig-$1$ to $12$ represent the phase portraits with respect to some suitable values of $ \beta $, $ \nu $, $ \alpha $, $ I_{S} $ and $ \Omega_{m} $.\\
\end{center}

In Table 4, for unstable critical points $ C_{5d} $ and $ C_{5e} $, we got values which are far outside of range, so we neglected this type of points in our further discussion. From above Table 4 to Table 9 (see Fig 1 to 6) we got a total of twelve stable points which are $ C_{5b} $, $ C_{5c} $, $ C_{6b} $, $ C_{6c} $, $ C_{7b} $, $ C_{7c} $, $ C_{8b} $, $ C_{8c} $, $ C_{9b} $, $ C_{9c} $, $ C_{10b} $ and $ C_{10c} $. For all of these points $-\frac{1}{3}<\omega_{eff}<-1$ and $q<0$. Again, we have stable critical points $ C_{11a} $, $ C_{11b} $, $ C_{12a} $, $ C_{12b} $, $ C_{13a} $, $ C_{13b} $, $ C_{14a} $ and $ C_{14b} $ related to Table 10 to Table 13 (see Fig 7 to 10). For these critical points the value of  $\omega_{eff}$ gradually decreases and tends to -1. From Table 14 and Table 15 (see Fig 11 and 12) we obtained stable critical points  $ C_{15a} $, $ C_{15b} $, $ C_{16a} $ and $ C_{16b} $. For these points $\omega_{eff}<-1$.

Amongst above critical points $ C_{5b} $, $ C_{6b} $, $ C_{7b} $, $ C_{8b} $, $ C_{9b} $, $ C_{10b} $, $ C_{11a} $, $ C_{12a} $, $ C_{13a} $, $ C_{14a} $, $ C_{15a} $ and $ C_{16a} $ describe the energy dominated era as well as the late time acceleration of our universe simultaneously. Though the stable critical points $ C_{5c} $, $ C_{6c} $, $ C_{7c} $, $ C_{8c} $, $ C_{9c} $, $ C_{10c} $, $ C_{11b} $, $ C_{12b} $, $ C_{13b} $, $ C_{14b} $, $ C_{15b} $ and $ C_{16b} $ signify both energy dominated era and late-time acceleration of our present universe, the solutions of these points are remain invalid as for these points $y < 0$ represents contracting universe which is contrary with our observations.

\section{Brief Discussion and Conclusions}

In our work the main motif is the interaction term $\beta$ which indicates the energy transfer between DE and DM. Now from Table 2, it is clear that when $\beta=0$ there is no existence of DE i.e., there is no energy transfer. When we observed the trace of DE in the DM dominated era the value of $\beta$ is very small ie, $\beta=0.001$. After that as the value of DE increases, $\beta$ also increases ie, the energy transfer between DM and DE increase due to increase in the value of DE.

In Table 2 $x=y=0$ signifies the flatness of the universe during above mentioned era. Except Table 2, for all other cases which indicate the late-time acceleration of the universe, we have $x\neq0$ and $y\geq0$. Again $y<0$ signifies the contracting universe.

When we entered in DE dominated era, From Table 3, we got large values of $\beta$ indicating more energy transfer between DE and DM. Again, when $\alpha=-5.05$, we got some stable points which signifies the DE dominated era of our universe. Even if we take $\alpha=-5.05$ , critical point $C_{3f}$ becomes unstable. Values of density parameters and interaction between DE and DM related with this critical point represents probable change in the values of $\Omega_d$ and $\Omega_m$ in future. It may be seen that the value of $\Omega_m$ will further increase in future whereas the value of $\Omega_d$ will decrease if there is more further interaction between DE and DM. For our three-dimensional interactive system, we did not get any viable solution to signify DE dominated era and late-time acceleration both at same time. We got some stable points $ C_{5b} $, $ C_{6b} $, $ C_{7b} $, $ C_{8b} $, $ C_{9b} $, $ C_{10b} $, $ C_{11a} $, $ C_{12a} $, $ C_{13a} $, $ C_{14a} $, $ C_{15a} $ and $ C_{16a} $ which represent DE dominated era as well as the late time acceleration of our ruling universe. Though the stable critical points  $ C_{5c} $, $ C_{6c} $, $ C_{7c} $, $ C_{8c} $, $ C_{9c} $, $ C_{10c} $, $ C_{11b} $, $ C_{12b} $, $ C_{13b} $, $ C_{14b} $, $ C_{15b} $ and $ C_{16b}$ also signify both the energy dominated era and late-time acceleration of the universe, the solutions of these points are remain invalid as for these points $y < 0$ represents the contracting universe which contradicts with our observations.

From figures, we have got variations in values of $\omega_{eff}$ and $q$. Applying these variations, we have also explained the different phases of acceleration of universe. We have $-\frac{1}{3}<\omega_{eff}<-1$ and $q<0$ for stable critical points of Table 4 to Table 13 (Fig. 1 to 10). These points signify quintessence field. Value of $\omega_{eff}$ is lesser than $-1$ for stable critical points in Table 13 (Fig. 10). This represents the phantom barrier ($\Lambda$CDM model). Again, for critical points in Table 14 and Table 15 (Fig. 11 and 12), we have obtained the scenario of $\omega_{eff}<-1$ and $q<0$. This signifies the universe will expand with acceleration beyond phantom barrier in future. Future deceleration is not supported so far if dark matter is interacting with dark energy. As dark matter is acting as an acute power supply in the conversion engine, dominance of dark energy will go on to give rise to a future cosmological singularity like Big Rip.

\vspace{.1 in}
{\bf Acknowledgment:}
This research is supported by the project grant of Government of West Bengal, Department of Higher Education, Science and Technology and Biotechnology (File no:- $ST/P/S\&T/16G-19/2017$). GD thanks Government of West Bengal, India for State Funded Fellowship. MB  thanks Department of Mathematics, The University of Burdwan for the research facilities provided. RB thanks IUCAA, Pune, India for Visiting Associateship.

\end{document}